\documentclass{article}
\usepackage[dvipdfmx]{graphicx}
\usepackage{
spconf,
amsmath,
amsfonts,
cite,
xurl,
color,
mathrsfs,
multirow,
array,
colortbl
}

\newcommand{\bm}[1]{\boldsymbol{#1}}

\newcommand{\mysection}[1]{
\vspace{-4pt}
\section{#1}
\vspace{-3pt}
}
\newcommand{\mysubsection}[1]{
\vspace{-5pt}
\subsection{#1}
\vspace{-2pt}
}

\newcommand{\mysubsubsection}[1]{
\vspace{-4pt}
\subsubsection{#1}
\vspace{-2pt}
}

\title{
Invertible {DNN}-based nonlinear time-frequency transform \\ for speech enhancement 
}

\name{Daiki Takeuchi$^\dagger$, Kohei Yatabe$^\dagger$, Yuma Koizumi$^\ddag$, Yasuhiro Oikawa$^\dagger$, Noboru Harada$^\ddag$
\vspace{-3pt}}
\address{
$^\dagger$Department of Intermedia Art and Science, Waseda University, Tokyo, Japan\\
$^\ddag$NTT Media Intelligence Laboratories, Tokyo, Japan
}

\begin{document}
\ninept
\maketitle

\begin{abstract}
We propose an end-to-end speech enhancement method with trainable time-frequency~(T-F) transform based on invertible deep neural network~(DNN).
The resent development of speech enhancement is brought by using DNN.
The ordinary DNN-based speech enhancement employs T-F transform, typically the short-time Fourier transform~(STFT), and estimates a T-F mask using DNN.
On the other hand, some methods have considered end-to-end networks which directly estimate the enhanced signals without T-F transform.
While end-to-end methods have shown promising results, they are black boxes and hard to understand.
Therefore, some end-to-end methods used a DNN to learn the linear T-F transform which is much easier to understand.
However, the learned transform may not have a property important for ordinary signal processing.
In this paper, as the important property of the T-F transform, perfect reconstruction is considered.
An invertible nonlinear T-F transform is constructed by DNNs and learned from data so that the obtained transform is perfectly reconstructing filterbank.
\end{abstract}


\mysection{Introduction}
Speech enhancement is used to recover the target speech from a noisy observed signal.
In the case of a single channel, the standard method is time-frequency~(T-F) masking which applies a mask in the T-F domain.
The performance of speech enhancement using T-F masking is affected by both T-F mask estimator and T-F transform.
The recent advance of T-F mask estimator is brought by DNN-based T-F mask estimation methods \cite{
wang2018supervised
,erdogan2015phase
,williamson2017time
,Zhao2018convolutional
,koizumi2018dnn
,fu2019metricgan
,Takeuchi2020real
,Koizumi2020Speech
,Kawanaka2020stable
}.
While DNN-based T-F masking is ordinarily applied in short-time Fourier transform~(STFT) domain, some methods designed a specific T-F transform for assisting T-F mask estimation and investigated optimal T-F domain for speech enhancement \cite{koizumi2019trainable,takeuchi2019data}.

Recently, some end-to-end speech enhancement methods which directly handle time-domain signals are proposed \cite{
pascual2017segan
,germain2018speech
,rethage2018wavenet
,baby2019sergan
,Pandey2019TCNN
}.
Among those speech enhancement methods, some methods have proposed DNN which plays the role of T-F transform and its inverse.
Since the end-to-end methods can obtain better T-F domain representation by learning from data, these methods outperformed speech enhancement methods performed in STFT domain.
This is because they can simultaneously train both T-F mask estimator and T-F transform.
However, it is hard to understand the role of each component in the trained DNN.
To understand the better T-F domain representation which is learned, the structure studied in signal processing is required.

The DNNs acting as T-F transform and its inverse are treated as analysis and synthesis filterbanks by introducing the structure of filterbank to DNNs.
Therefore, the theory of filterbank can be applied to DNN.
Considering as filterbank, the synthesis part is important for reconstructing the original time-domain signals.
The property of reconstructing the original signal is called perfect reconstruction which indicates the invertible transform.
The perfect reconstruction property is important, and it should be investigated.

\begin{figure}[t]
  \centering
  \includegraphics[width=0.95\columnwidth]{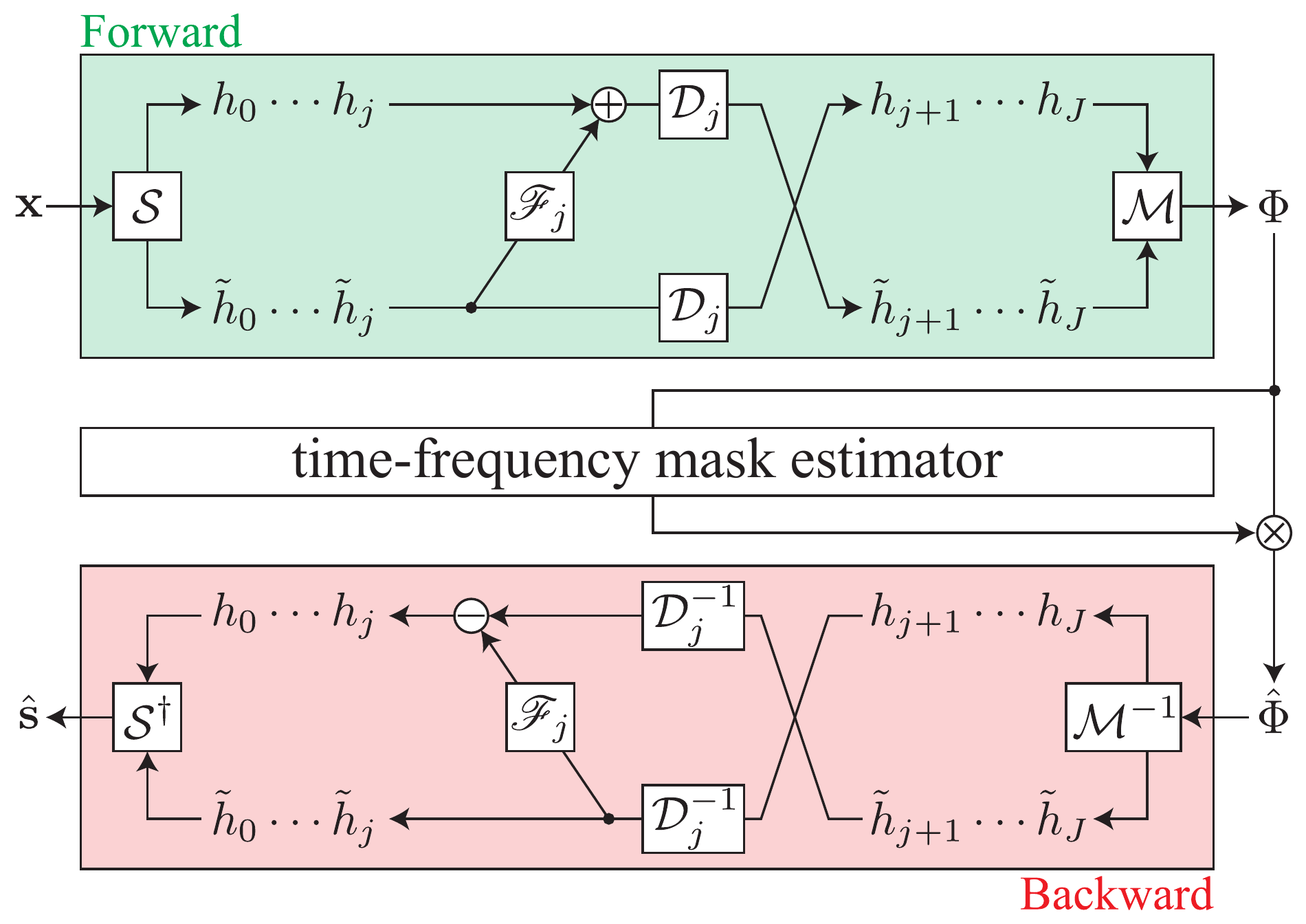}
  \vspace{-8pt}
  \caption{
  Illustration of the structure of the proposed method.  
  $j = 1,\ldots, J$, $\mathscr{F}_j$, $\mathcal{S}$, $\mathcal{D}_j$, and $\mathcal{M}$ are lifting indices, $j$th DNN block, splitting operator, $j$th invertible down sampling, and merging operator, respectively. $\cdot^{-1}$ and $\cdot^{\dagger}$ denote inverse and generalized inverse.
  The transformed feature without masking $\Phi$ can be perfectly transformed back to input $\mathbf{x}$ by the backward network.
  }
  \label{fig:iRevNetArch}
\vspace{2pt}
\end{figure}

In this paper, we propose an end-to-end speech enhancement method with a trainable T-F transform based on the invertible network.
As invertible DNN playing the role of T-F transform, the i-RevNet~\cite{jacobsen2018revnet} illustrated in Fig.~\ref{fig:iRevNetArch} is used among the other invertible DNNs\cite{
dinh2016density
,kingma2018glow
,ho2019flow++
,behrmann2019invertible
,prenger2019waveglow
}.
The i-RevNet has the forward block and the backward block, and each block consists of the inverse function of each other. 
Since the i-RevNet is always invertible because of the structure, a cost function or learning method to guarantee the invertibility is not required unlike \cite{prenger2019waveglow}.
A T-F mask is applied in the T-F domain learned by the i-RevNet for enhancing the speech signals.
According to experimental results, speech enhancement can be achieved by only learning T-F transform without learning T-F mask estimator.

\mysection{Preliminaries}

\vspace{6pt}
\mysubsection{DNN-based speech enhancement}
The problem of speech enhancement is to recover a target speech signal $\mathbf{s}~\in~\mathbb{R}^T$ degraded by noise $\mathbf{n}$.
An observed signal is modeled as 
\begin{equation}
    \mathbf{x} = \mathbf{s} +\mathbf{n}.
\end{equation}
In DNN-based speech enhancement with T-F masking, the estimated speech signal $\hat{\mathbf{s}}$ is given as
\begin{equation}
\hat{\mathbf{s}} = \mathcal{F}^{\dagger}( \mathcal{M}_\theta (\Psi) \odot \mathcal{F}(\mathbf{x}) ),
\end{equation}
where $\mathcal{F}$ is T-F transform, $\mathcal{M}_\theta$ is a regression function implemented by DNN, $\theta$ is a set of its parameters, $\Psi$ is the input acoustic feature, $\cdot^\dagger$ denotes generalized inverse, and $\odot$ denotes element-wise multiplication, respectively.
In many methods of DNN-based speech enhancement, STFT is used as the T-F transform $\mathcal{F}$. 
Since inverse STFT can be designed to reconstruct data from T-F domain perfectly, no information loss happens by the transformation.
This perfect reconstruction property is the important ingredient of speech enhancement because the enhanced result must be converted back into the time domain after applying a T-F mask in the T-F domain.

\mysubsection{End-to-end speech enhancement}
While DNN-based T-F masking in STFT domain performs well for speech enhancement, the end-to-end speech enhancement method has outperformed those T-F-masking-based methods in STFT domain~\cite{pascual2017segan
,germain2018speech
,rethage2018wavenet
,baby2019sergan}.
Since DNN is applied as a function from time domain signal to time domain signal in the end-to-end method as
\begin{equation}
    \hat{\mathbf{s}} = \mathscr{D}_\phi(\mathbf{x}),
\end{equation}
it is hard to understand how the DNN $\mathscr{D}$ enhances speech in those method.
To interpret the end-to-end speech enhancement from the viewpoint of signal processing, some insight from signal processing should be introduced to DNN.

Among the end-to-end methods for speech enhancement, some methods used DNN which plays the role of T-F transform.
These methods transform a time-domain signal to some T-F domain constructed by DNN.
Then, a signal in T-F domain is filtered by a T-F mask estimated by another DNN and transformed back to the time domain by the other DNN acting as the inverse T-F transform.
This process of obtaining the enhanced signal $\hat{\mathbf{s}}$ can be written as
\begin{equation}
\hat{\mathbf{s}} = \mathcal{P}^{\dagger}_{\beta}(
\mathcal{M}_{\theta}(\Psi) \odot \mathcal{P}_{\alpha}(\mathbf{x})
),
\end{equation}
where $\mathcal{P}$ and $\mathcal{P}^{\dagger}$ are the DNNs acting as the T-F transform and its inverse transform, $\alpha$ and $\beta$ represent the sets of parameters of $\mathcal{P}$ and $\mathcal{P}^{\dagger}$, and $\odot$ denotes element-wise multiplication.
The parameters $\alpha$ and $\beta$ are learned from data through the training. 
Namely, the filterbanks $\mathcal{P}$ and $\mathcal{P}^{\dagger}$ are obtained by training the DNNs.
Therefore, it is assumed that the theory of filterbanks can be used for the end-to-end speech enhancement by introducing the property of the filterbank to DNN.
In the area of filterbank, a filterbank having the inverse transform is called as the perfect reconstruction filterbank, which is well studied in signal processing.
In general, a pair of learned DNNs $\mathcal{P}$ and $\mathcal{P}^{\dagger}$ for speech enhancement cannot reconstruct the original signal as $\mathbf{s} = \mathcal{P}^{\dagger}_{\beta}(\mathcal{P}_{\alpha}(\mathbf{s}))$ without a special treatment.
DNNs should be designed as perfect reconstruction filterbanks so that a signal can be transformed back to the time domain.

\begin{figure}[t]
  \centering
  \includegraphics[width=0.99\columnwidth]{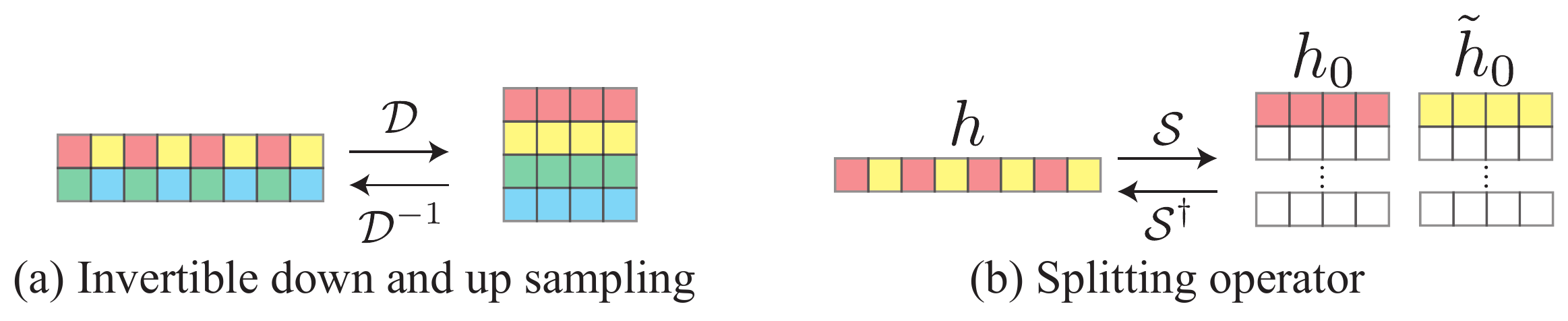}
  \vspace{-9pt}
  \caption{(a)~Illustration of the invertible down and up sampling in the i-RevNet. (b)~Illustration of the splitting operator $\mathcal{S}$.}
  \label{fig:invDS}
  \vspace{0pt}
\end{figure}

\mysubsection{Invertible deep neural network}
Recently, the invertible DNNs, which have their inverse functions, are studied and applied to various tasks including the generative model of image and the speech synthesis\cite{dinh2016density,kingma2018glow,ho2019flow++,behrmann2019invertible,prenger2019waveglow}.
In these methods, invertibility of DNN is imposed by the structure of DNN.
The invertible structure can be divided into two types: the structures which may have the inverse and the structures which always have the inverse.
One of the former structures is the invertible $1 \times 1$ convolutional layer\cite{kingma2018glow}.
Since the structure of the invertible $1 \times 1$ convolutional layer is almost the same as the standard $1 \times 1$ convolutional layer, there is no disadvantage by imposing the invertibility.
However, the cost function for training is required to keep invertibility apart from the cost function for solving the task because that structure does not guarantee the invertibility.
As the structures which always have inverse, the affine coupling layer\cite{dinh2016density, kingma2018glow} is often used in invertible DNNs.
Since the inverse of the affine coupling layer always exists, no cost function is required to keep the invertibility.
However, the expressive power might be reduced because one affine coupling layer only transforms a half of signal in channel dimension.

\mysection{Proposed method}

As discussed in the previous section, it is desired that the DNN-based T-F transform has the perfect reconstruction property.
In this paper, we proposed to use the i-RevNet\cite{jacobsen2018revnet}, which is one of invertible neural network, as T-F transform in DNN-based speech enhancement as in Fig.~\ref{fig:iRevNetArch}.
The i-RevNet consists of the forward network and the backward network.
In the proposed method, the forward network of i-RevNet is used as T-F transform, and the backward network is used as its inverse transform.
From invertibility of the i-RevNet, the proposed T-F transform always has the perfect reconstruction property.

\begin{figure*}[t]
  \centering
  \includegraphics[width=1.7\columnwidth]{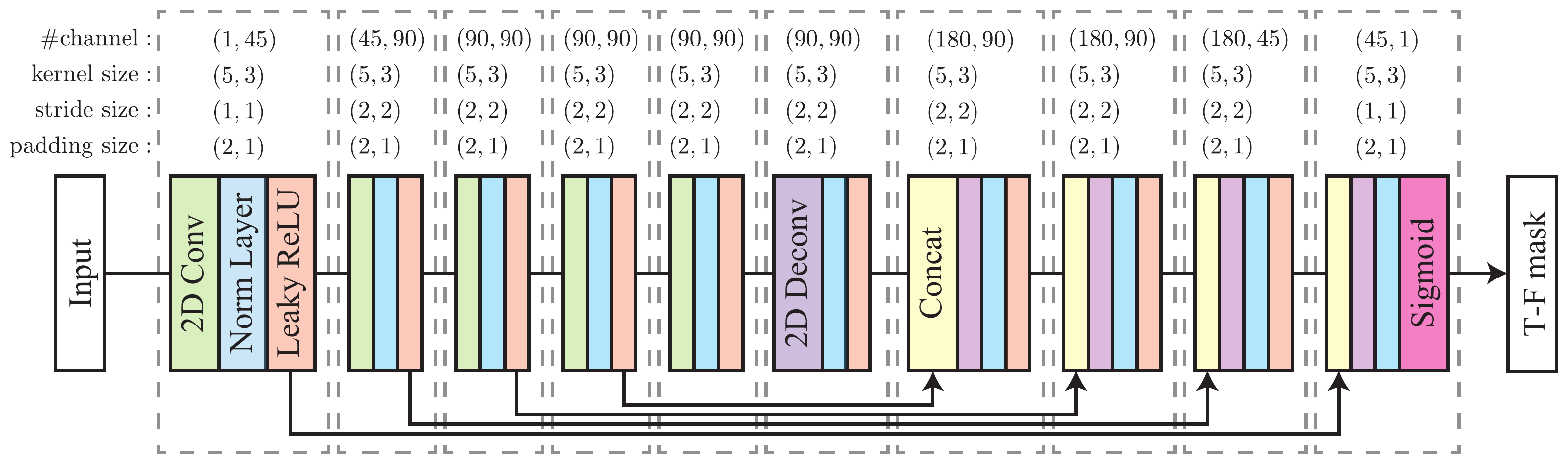}
  \vspace{-0pt}
  \caption{Illustration of DNN used in the experiment. 
  ``Norm Layer", ``2D Conv", ``2D Deconv" and ``Concat" stand for
  normalization layer, two dimensional convolution, two dimensiona deconvolution, and concatenation, respectively.
  Normalization layer is spectral normalization\cite{miyato2018spectral} or instance normalization.}
  \label{fig:UNetArch}
\end{figure*}

\mysubsection{i-RevNet as T-F transform}
The i-RevNet is one of the invertible neural networks illustrated in Fig.~\ref{fig:iRevNetArch}.
To introduce DNN while maintaining invertibility, the affine coupling layer which is inspired by the lifting scheme of wavelet transform is repeatedly used in the i-RevNet.
Invertibility of the affine coupling layer is preserved no matter what function with any nonlinearity is used for $\mathscr{F}$ in Fig.~\ref{fig:iRevNetArch}.
Thus, the use of i-RevNet as T-F transform enables to obtain the trainable nonlinear T-F transform which has perfect reconstruction property and can be trained by back-propagation.  
In the proposed method, each DNN block $\mathscr{F}_j$ consists of 1D-convalitional layers.
Note that, both of forward and backward network of the i-RevNet use common DNN blocks $\mathscr{F}_j$. 
That is, the parameters of DNNs acting as T-F transform $\mathcal{P}$ and its inverse $\mathcal{P}^{\dagger}$ are the same.
The invertible down sampling $\mathcal{D}_j$ in the i-RevNet uses reshaping instead of dilating as illustrated in Fig.~\ref{fig:invDS}(a).
The splitting operator $\mathcal{S}$ divides a time-domain signal into odd and even components and increases the channel dimensionality by concatenating $0$ as in Fig.~\ref{fig:invDS}(b).
The merging operator $\mathcal{M}$ only concatenates $x_J$ and $\tilde{x}_J$ in channel dimension, and its inverse separates the feature $\Phi$ into $x_J$ and $\tilde{x}_J$.

\mysubsection{Proposed end-to-end speech enhancement method}
We propose the end-to-end speech enhancement method using the i-RevNet instead of the ordinary T-F transform. 
In the proposed method, i-RevNet is used as T-F transform, and a T-F mask is estimated in the T-F domain generated by i-RevNet.
The input signal in time domain is transformed to the signal in T-F domain by the forward network of i-RevNet.
After multiplication of the input signal and a T-F mask in the T-F domain, the enhanced signal in the time domain is calculated by the backward network of i-RevNet.
In the step of estimating the T-F mask, the channel dimension of the output of i-RevNet is treated as the height dimension for applying two dimensional convolution.
Our implementation used in the following experiment is openly available on web\footnote{\url{https://github.com/dtake1336/i-revnet-based-time-frequency-transform}}.

\mysection{Experiment}

\vspace{6pt}
\mysubsection{Experimental condition}

\vspace{6pt}
\mysubsubsection{Dataset}
We utilized the VoiceBank-DEMAND dataset constructed by Valentini 
{\it et al}.~\cite{valentini2016investigating} which is openly available%
\footnote{\url{http://dx.doi.org/10.7488/ds/1356}} and frequently used in experiments of DNN-based speech enhancement. 
It consists of train set and test set which contain noisy mixtures and clean speech signals, respectively, i.e., noise and speech signals were already mixed by the authors \cite{valentini2016investigating}.
The train and test sets consist of 28 and 2 speakers (11\,572 and 824 utterances) \cite{veaux2013voice}, respectively, which are contaminated by 10 (DEMAND, speech-shaped noise, and babble) and 5 types of noise (DEMAND) \cite{thiemann2013diverse}, respectively.
All data were downsampled from 48 kHz to 16 kHz.

\begin{table*}[t]
\centering
\caption{Results of experiment}
\vspace{2pt}
\begin{tabular}{ c | c | c | c | c | c | c | c  } \hline \hline
T-F transform & DNN block $\mathscr{F}$ (Activation) &T-F mask (Normalization) & SI-SDR imp. & PESQ & CSIG & CBAK & COVL \\ \hline
i-RevNet & U-Net (leaky ReLU)  & binary(N/A) & $\bm{9.79}$ & 2.48 & 3.49 & 2.60 & 2.96 \\ \hline
i-RevNet  & No Bias U-Net (N/A) & binary(N/A) & 7.00 & 2.12 & 3.12 & 2.37 & 2.57 \\ \hline

i-RevNet  & U-Net (leaky ReLU)  & U-Net (SN) & 9.54 & 2.49 & 3.55 & 2.61 & 3.00 \\ \hline
i-RevNet  & No Bias U-Net (N/A) & U-Net (SN) & 9.00 & 2.34 & 3.33 & 2.53 & 2.82 \\ \hline

i-RevNet  & U-Net (leaky ReLU)  & U-Net (IN) & 9.28 & 2.33 & 3.28 & 2.53 & 2.79 \\ \hline
i-RevNet  & No Bias U-Net (N/A) & U-Net (IN) & 8.97 & 2.49 & $\bm{3.65}$ & 2.60 & $\bm{3.04}$ \\ \hline

STFT  & N/A & U-Net (SN) & 8.52 & $\bm{2.54}$ & 3.52 & $\bm{2.62}$ & 3.01 \\ \hline
STFT  & N/A & U-Net (IN) & 8.66 & $\bm{2.54}$ & 3.57 & $\bm{2.62}$ & $\bm{3.04}$ \\ \hline
\hline
\end{tabular}
\label{tab:Results}
\vspace{-2pt}
\end{table*}

\mysubsubsection{DNN architecture, loss function and training setup}
In the experiment, the architecture illustrated in Fig.~\ref{fig:iRevNetArch} was used for the proposed method.
The number of lifting $J$ was set to 6 in all experiments in this section.
 
\begin{figure}[t]
  \centering
  \includegraphics[width=0.98\columnwidth]{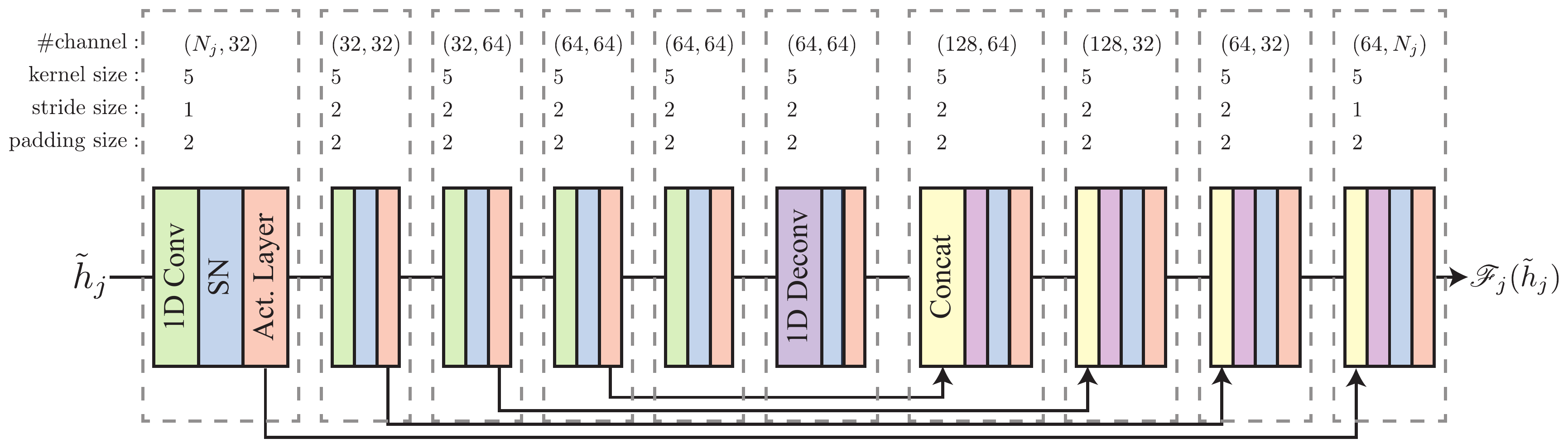}
  \vspace{-2pt}
  \caption{Illustration of DNN block $\mathscr{F}_j$ of the i-RevNet. ``SN", ``1D Conv", and ``1D Deconv" stand for spectral normalization one dimensional convolution, and one dimensional deconvolution, respectively. }
  \label{fig:iRevNetBlock}
  \vspace{-5pt}
\end{figure}

In the splitting operator, channel dimensionality was increased so that the number of elements of feature $\Phi$ was four times that in the input signal $h$.
DNN block $\mathscr{F}_j$ was 1D-convolutional-layer-based CNN summarized in Fig.~\ref{fig:iRevNetBlock}.
Since the number of channels in each i-RevNet layer $N_j$ was set to $4 \cdot 2^{j-1}$, the size of the T-F domain signal obtained from the time domain signal $\mathbf{x} \in \mathbb{R}^T$ was $256 \times (T/64)$.
In order to investigate the effect of introducing nonlinearlity to the T-F transform, the i-RevNet whose DNN block excludes leaky ReLU layer and bias of 1D convolutional layer was also used.

\begin{figure}[t]
  \centering
  \includegraphics[width=0.93\columnwidth]{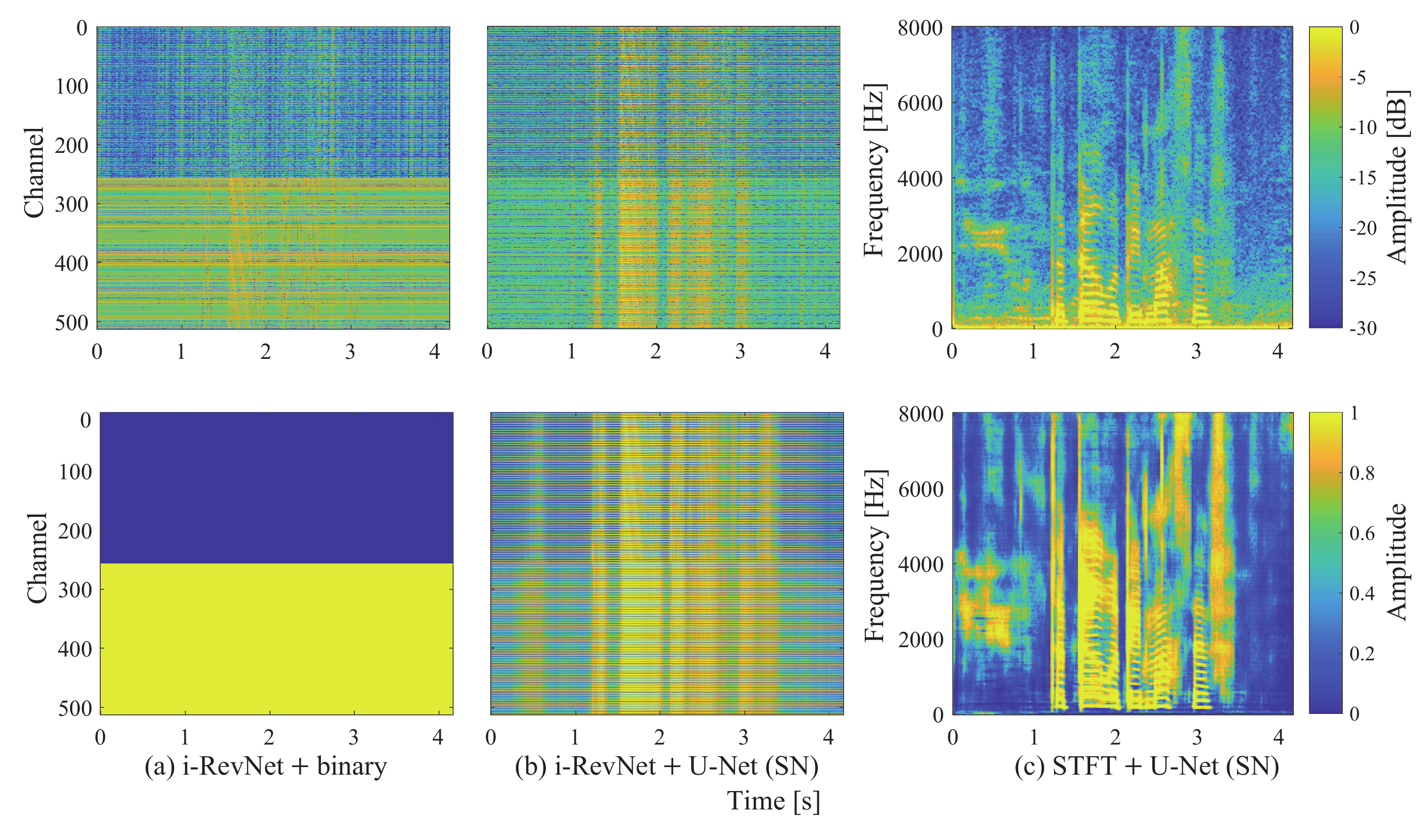}
  \vspace{-9pt}
  \caption{T-F representation (upper) and T-F mask (lower). In the proposed method (a), T-F mask is fixed, so that T-F transform should learn to discriminate speech and noise and assign them to the corresponding channels.}
  \label{fig:tfRepComp}
  \vspace{-3pt}
\end{figure}

In the T-F masking step, the discriminative binary mask and DNN-estimated mask were used.
All elements of the discriminative binary mask entries are 0 or 1 shown in Fig.~\ref{fig:tfRepComp}, and there is a no fluctuation in time dimension corresponding to voice activity. Thus, T-F transform is required to discriminate speech and noise and assign them to the corresponding channels.
In DNN-based mask estimation, U-Net-like architecture illustrated in Fig.~\ref{fig:UNetArch} excluding the feature extraction layer was applied as T-F mask estimator.

For the loss function in training, SDR-based loss was used:
\begin{align}
\mathcal{J}_{\rm SDR}(\theta) 
&= \frac{1}{2}(
{\rm clip}_\beta[{\rm SDR}(\hat{\mathbf{s}}, \mathbf{s})]
+
{\rm clip}_\beta[{\rm SDR}(\mathbf{x}-\hat{\mathbf{s}}, \mathbf{n})]
)
,\label{eq:SDRloss}     
\end{align}
where ${\rm SDR}(\mathbf{s},\mathbf{y}) = 10 \log_{10}(\|\mathbf{s}\|_2^2 / \|\mathbf{s}-\mathbf{y}\|_2^2)$, $\|\cdot\|_2^2$ is $\ell_2$ norm, ${\rm clip}_\beta[\mathbf{x}] = \beta \cdot {\rm tanh}(x/\beta)$, and $\beta>0$ is a clipping parameter\cite{Erdogan2018investigations}.

As the conventional method, DNN-based speech enhancement in STFT domain was considered.
STFT with the 512 points (32 ms) Hann window, 128 points time-shifting and 512 points discrete Fourier transform length was used, and the inverse STFT was implemented by its canonical dual \cite{yatabe2019DGT} to make the STFT as perfect reconstruction filterbank. 
To estimate real-valued T-F mask, U-Net illustrated in Fig.~\ref{fig:UNetArch} was used.
The log-magnitude spectrogram was used as the input feature:
\begin{equation}
\Psi = {\rm ln}(|{\rm STFT}(\mathbf{x})|),
\end{equation}
where ${\rm STFT}$ denotes STFT operator.
As an activation function of the output layer, the sigmoid function was used for limiting the values within the range 0 to 1.
The loss function used for the proposed method, Eq.~\eqref{eq:SDRloss}, was also used for the conventional method.

DNN in the proposed and conventional methods were trained 500 epochs.
Since 10 percent of the train set is used for validation, 10 415 utterances are used for training. 
Mini-batch size was 16 and Adam\cite{kingma2015adam} whose learning rate was fixed to 0.0001 was utilized as the optimizer for training DNN.
The performance of speech enhancement was measured by SI-SDR~\cite{le2019sdr}, PESQ\cite{wPESQ}, and three measures CSIG, CBAK, and COVL~\cite{Hu2008eval} which are the popular predictor of the mean opinion score~(MOS) of the signal distortion, the background noise interference, and the overall effect, respectively.
SI-SDR is given by
\begin{align}
{\rm SI \mathchar`- SDR} &= 10\log_{10}
\frac{
\| \gamma \mathbf{s} \|^2_2
}{
\| \gamma \mathbf{s} - \hat{\mathbf{s}} \|^2_2
}
,
\end{align}
where $\gamma = 
\mathbf{s}^\mathsf{T} \hat{\mathbf{s}}
/
\| \mathbf{s} \|^2_2
$ and $\cdot^\mathsf{T}$ denotes transpose.

\mysubsection{Results}

\begin{figure}[t]
  \centering
  \includegraphics[width=0.93\columnwidth]{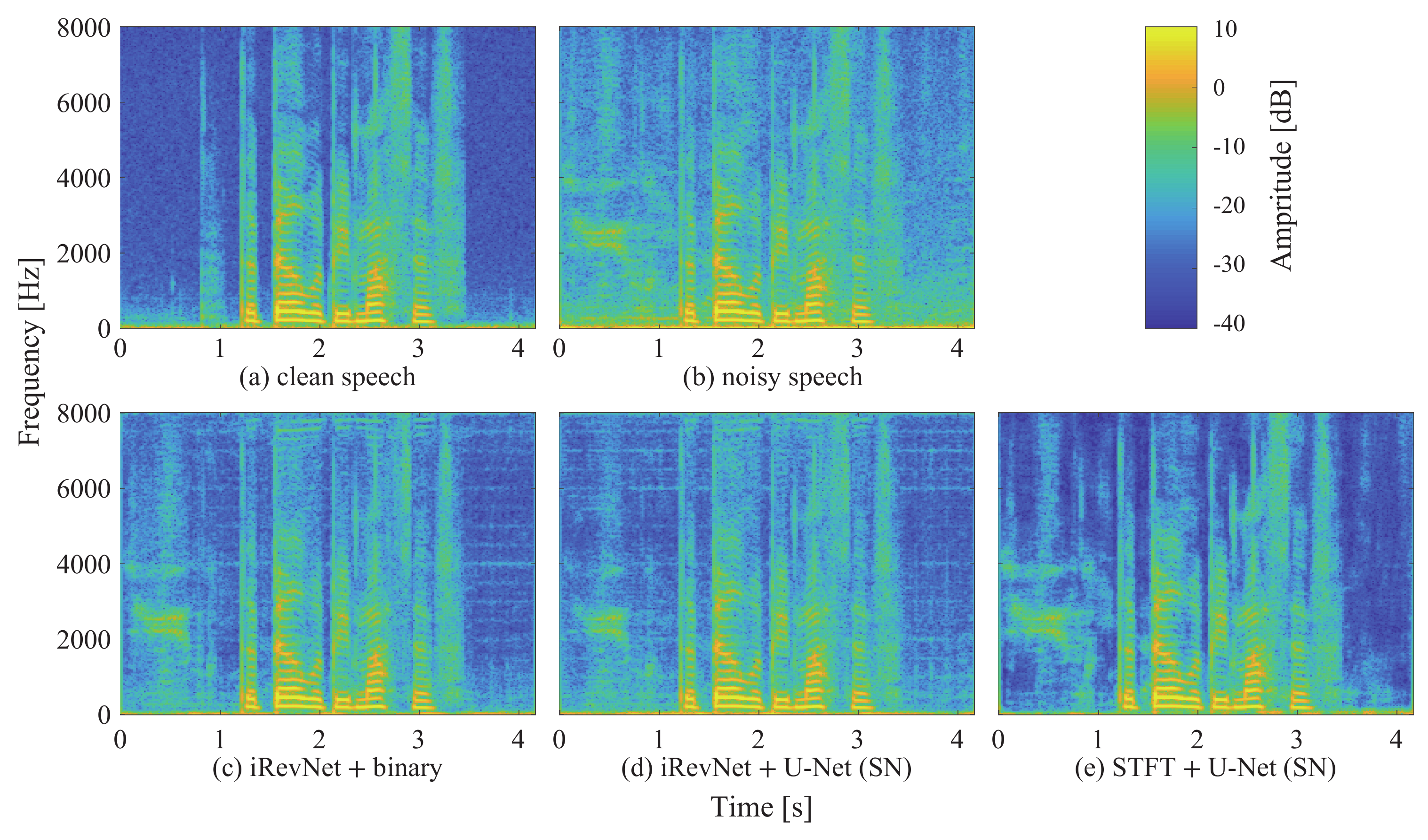}
  \vspace{-9pt}
  \caption{Example of spectrogram enhanced by the proposed and conventional methods. Two figures in upper row are clean and noisy speeches.
  Three figures in lower row are enhanced speech obtained by the proposed and conventional methods.}
  \label{fig:enhSpec}
  \vspace{-4pt}
\end{figure}

The results are summarized in Table~\ref{tab:Results}.
``DNN block $\mathscr{F}$'' represents the architecture of the DNN block of the i-RevNet.
``U-Net'' is shown in Fig~\ref{fig:iRevNetBlock}, and ``No Bias U-Net'' excludes the bias of each 1D convolutional layer.
``U-Net'' in ``T-F mask'' means that the T-F mask was estimated by the U-Net shown in Fig.~\ref{fig:UNetArch}, and ``binary'' is the discriminative binary mask shown in Fig.~\ref{fig:tfRepComp}(a).

The method using i-RevNet as T-F transform obtained comparable score to the conventional method which uses U-Net as T-F mask estimator in STFT domain.
Since the best score of SI-SDR is obtained by i-RevNet with the discriminative binary mask, speech enhancement can achieve with only learning T-F transform without learning DNN-based T-F mask estimator.
In the case of the DNN-estimated mask, PESQ, CSIG, CBAK, and COVL were improved compared to the case of the discriminate binary mask.
Since the DNN-estimated mask was learned the fluctuation in time dimension like voice activity, these perceptual evaluation measures should be improved.

The spectrograms of the speech enhanced by the proposed and conventional methods (first, third and seventh rows in Table~\ref{tab:Results}) are shown in Fig.~\ref{fig:enhSpec}~(c), (d), and (e).
The spectrograms of the proposed method in Figs.~\ref{fig:enhSpec}~(c) and (d) have some horizontal pattern and aliasing in the enhanced speech.
Since the proposed architecture of T-F transform has the dilation in the splitting operator and invertible down sampling operator, the time domain signal processed through them may contain specific tendency in the frequency direction.

When focusing on the presence of the activation function and the bias of 1D convolutional layer in the DNN block of i-RevNet, the methods with nonlinear activation functions and the bias obtained the higher SI-SDR improvement than the other without nonlinearlity.
In the case of DNN-estimated mask, the difference of measures between the method with nonlinear activation functions and one without nonlinearlity decreased compared to the case of the discriminative binary mask, and the highest CSIG is obtained by the nonlinear i-RevNet.
Therefore, the expressive power of the i-RevNet as T-F transform can be improved by introducing the nonlinear functions.
Meanwhile, the nonlinearlity of T-F transform and one of T-F mask estimator may conflict so that there are some inconveniences for training.
From these results, it is confirmed that the trainable and nonlinear T-F transform for speech enhancement can be designed by the use of the i-RevNet.

\vspace{-3pt}
\mysection{Conclusion}
\vspace{-2pt}
In this paper, an end-to-end speech enhancement method with trainable T-F transform based on invertible DNN is proposed. 
By the use of i-RevNet as T-F transform, trainable T-F transform which has perfect reconstruction property is realized. 
Since i-RevNet is invertible without constraint in training, the proposed T-F transform can be learned by only the cost function for speech enhancement.
Future works include analysis of the learned T-F transform to investigate the optimal T-F transform for speech enhancement.


\end{document}